# H Index of scientific Nursing journals according to Google Scholar Metrics (2007-2011)

Liliana Marcela Reina Leal[1], Rafael Repiso[2], Emilio Delgado López-Cózar[3]

[1] RN, Grupo de Estudios Documentales - Fundación Index. Granada, Spain. Master student, Máster en Información y Comunicación Científica, Universidad de Granada

[2,3] EC3 Research Group & EC3metrics Spin-Off, Universidad de Granada, Granada (Spain)

**ABSTRACT**

The aim of this report is to present a ranking of Nursing journals covered in Google Scholar Metrics (GSM), a Google product launched in 2012 to assess the impact of scientific journals from citation counts this receive on Google Scholar. Google has chosen to include only those journals that have published at least 100 papers and have at least one citation in a period of five years (2007-2011). Journal rankings are sorted by languages (showing the 100 papers with the greatest impact). This tool allows to sort by subject areas and disciplines, but only in the case of journals in English. In this case, it only shows the 20 journals with the highest h index. This option is not available for journals in the other nine languages present in Google (Chinese, Portuguese, German, Spanish, French, Korean, Japanese, Dutch and Italian).

Google Scholar Metrics doesn't currently allow to group and sort all journals belonging to a scientific discipline. In the case of Nursing, in the ten listings displayed by GSM we can only locate 34 journals. Therefore, in an attempt to overcome this limitation, we have used the diversity of search procedures allowed by GSM to identify the greatest number of scientific journals of Nursing with h index calculated by this bibliometric tool. Bibliographic searches were conducted between 10th and 30th May 2013.

The result is a ranking of 337 nursing journals sorted by the same h index, and mean as discriminating value. Journals are also grouped by quartiles.

## METHODOLOGICAL NOTE

- Subject area covered: Nursing journals included in Google Scholar Metrics.
- Google Scholar Metrics is the data source used, whose coverage is at least 100 journals published during the 2007-2011 period, and with at least one citation.
- The journals are sorted by their h index. In case of equality of it, discriminator value is the mean of the number of citations obtained by the articles that contribute to the h index.
- Searches were conducted between 10th and 30th May 2013.

**Ranking of Nursing journals covered in Google Scholar Metrics (2007-2011)**

| Rank | Quartil | | Journal Title | H Index | M$_{ed}$ H Index |
|---|---|---|---|---|---|
| 1 | Q1 |  | Journal of Advanced Nursing | 50 | 61 |
| 2 | Q1 |  | 中华护理杂志 (Chinese Journal of Nursing) | 46 | 66 |
| 3 | Q1 |  | International Journal of Nursing Studies | 43 | 54 |
| 4 | Q1 |  | Journal of Clinical Nursing | 43 | 53 |
| 5 | Q1 |  | Journal of Nursing Management | 34 | 44 |
| 6 | Q1 |  | 护理研究 (Nursing Research) | 32 | 40 |
| 7 | Q1 |  | Oncology Nursing Forum | 31 | 46 |
| 8 | Q1 |  | American Journal of Critical Care | 31 | 43 |
| 9 | Q1 |  | Journal of Nursing Administration | 30 | 40 |
| 10 | Q1 |  | Nurse Education Today | 30 | 38 |
| 11 | Q1 |  | Revista da Escola de Enfermagem da USP | 29 | 44 |
| 12 | Q1 |  | 中国护理管理 (Chinese Nursing Management) | 29 | 43 |
| 13 | Q1 |  | Birth | 28 | 43 |
| 14 | Q1 |  | Texto & Contexto-Enfermagem | 28 | 37 |
| 15 | Q1 |  | Revista Brasileira de Enfermagem | 28 | 35 |
| 16 | Q1 |  | Journal of Nursing Scholarship | 27 | 43 |
| 17 | Q1 |  | Nursing Research | 27 | 41 |
| 18 | Q1 |  | Journal of Obstetric, Gynecologic, and Neonatal Nursing | 27 | 38 |
| 19 | Q1 |  | Journal of Nursing Education | 27 | 37 |
| 20 | Q1 |  | Cancer Nursing | 26 | 33 |
| 21 | Q1 |  | Scandinavian Journal of Caring Sciences | 26 | 32 |
| 22 | Q1 |  | Journal of Midwifery & Women's Health | 26 | 31 |
| 22 | Q1 |  | 护理学杂志: 外科版 (Journal of Nursing Science: Surgical) | 26 | 31 |
| 24 | Q1 |  | Research in Nursing & Health | 25 | 42 |
| 25 | Q1 |  | Nursing Education Perspectives | 25 | 35 |
| 26 | Q1 |  | European Journal of Oncology Nursing | 25 | 34 |
| 27 | Q1 |  | Nursing Economics | 24 | 38 |
| 28 | Q1 |  | European Journal of Cardiovascular Nursing | 24 | 33 |
| 29 | Q1 |  | Revista Latino-Americana de Enfermagem | 24 | 30 |
| 30 | Q1 |  | Nursing Ethics | 24 | 29 |
| 31 | Q1 |  | Journal of Psychiatric & Mental Health Nursing | 24 | 28 |
| 31 | Q1 |  | Midwifery | 24 | 28 |
| 33 | Q1 |  | American Journal of Nursing | 23 | 37 |
| 34 | Q1 |  | Journal of Cardiovascular Nursing | 23 | 35 |
| 34 | Q1 |  | Journal of the American Academy of Nurse Practitioners | 23 | 35 |
| 36 | Q1 |  | Clinical Journal of Oncology Nursing | 23 | 33 |

| Rank | Q | Country | Journal | H | H5 |
|---|---|---|---|---|---|
| 36 | Q1 | UK | Journal of Wound Care | 23 | 33 |
| 38 | Q1 | Brazil | Acta Paulista de Enfermagem | 23 | 29 |
| 39 | Q1 | China | 中国实用护理杂志 (Chinese Journal of Practical Nursing) | 23 | 28 |
| 39 | Q1 | China | 护理管理杂志 (Journal of Nursing Administration) | 23 | 28 |
| 41 | Q1 | USA | Journal of Wound, Ostomy & Continence Nursing | 22 | 33 |
| 42 | Q1 | UK | European Journal of Cancer Care | 22 | 31 |
| 43 | Q1 | UK | Intensive and Critical Care Nursing | 22 | 30 |
| 43 | Q1 | China | 解放军护理杂志 (Journal of Nursing) | 22 | 30 |
| 43 | Q1 | USA | Nursing Outlook | 22 | 30 |
| 46 | Q1 | USA | Heart & Lung | 22 | 29 |
| 47 | Q1 | UK | Nurse Education in Practice | 22 | 27 |
| 48 | Q1 | USA | Western Journal of Nursing Research | 21 | 32 |
| 49 | Q1 | China | 护理学报 (Journal of Nursing) | 21 | 28 |
| 50 | Q1 | USA | Journal of Continuing Education in Nursing | 20 | 31 |
| 50 | Q1 | USA | Seminars in Oncology Nursing | 20 | 31 |
| 52 | Q1 | USA | Advances in Skin & Wound Care | 20 | 28 |
| 52 | Q1 | UK | British Journal of Nursing | 20 | 28 |
| 52 | Q1 | USA | Journal of Transcultural Nursing | 20 | 28 |
| 55 | Q1 | USA | Critical Care Nurse | 20 | 27 |
| 55 | Q1 | Germany | International Journal of Nursing Education Scholarship | 20 | 27 |
| 55 | Q1 | UK | International Nursing Review | 20 | 27 |
| 58 | Q1 | Australia | Australian Journal of Rural Health | 20 | 25 |
| 59 | Q1 | China | 中华现代护理杂志 (Modern Nursing) | 20 | 24 |
| 60 | Q1 | USA | Worldviews on Evidence-Based Nursing | 19 | 35 |
| 61 | Q1 | UK | Nursing in Critical Care | 19 | 31 |
| 62 | Q1 | USA | Journal of Pediatric Nursing | 19 | 29 |
| 63 | Q1 | USA | Journal of Professional Nursing | 19 | 28 |
| 64 | Q1 | USA | Journal of Emergency Nursing | 19 | 26 |
| 64 | Q1 | USA | Public Health Nursing | 19 | 26 |
| 66 | Q1 | Australia | International Journal of Mental Health Nursing | 19 | 25 |
| 66 | Q1 | Australia | International Journal of Nursing Practice | 19 | 25 |
| 68 | Q1 | USA | AORN Journal | 19 | 24 |
| 68 | Q1 | UK | Nursing Inquiry | 19 | 24 |
| 70 | Q1 | USA | Advances in Nursing Science | 18 | 29 |
| 71 | Q1 | USA | Nursing Forum | 18 | 27 |
| 72 | Q1 | Australia | Nursing and Health Sciences | 18 | 26 |
| 73 | Q1 | USA | Journal of Pediatric Healthcare | 18 | 25 |
| 74 | Q1 | USA | Association of Nurses in AIDS Care. Journal | 18 | 24 |
| 74 | Q1 | USA | Biological Research for Nursing | 18 | 24 |
| 74 | Q1 | USA | Journal of Gerontological Nursing | 18 | 24 |
| 74 | Q1 | USA | Journal of Nursing Care Quality | 18 | 24 |
| 78 | Q1 | USA | Journal of Human Lactation | 18 | 21 |
| 79 | Q1 | China | 护理与康复 (Nursing and Rehabilitation) | 17 | 24 |
| 80 | Q1 | USA | Journal of Pediatric Oncology Nursing | 17 | 23 |
| 80 | Q1 | USA | MCN: the American Journal of Maternal Child Nursing | 17 | 23 |
| 82 | Q1 | USA | Health Care for Women International | 17 | 22 |
| 82 | Q1 | USA | Issues in Mental Health Nursing | 17 | 22 |
| 82 | Q1 | UK | Nursing Standard | 17 | 22 |
| 82 | Q1 | China | 护理实践与研究 (Practice and Research) | 17 | 22 |
| 86 | Q2 | Brazil | Escola Anna Nery Revista de Enfermagem | 17 | 19 |
| 87 | Q2 | USA | Nursing Administration Quarterly | 16 | 24 |

EC3 Reports № 5| Rank | Quartile | Country | Journal | Col1 | Col2 |
|---|---|---|---|---|---|
| 88 | Q2 | 🇺🇸 | Policy, Politics & Nursing Practice | 16 | 22 |
| 89 | Q2 | 🇦🇺 | Contemporary Nurse | 16 | 21 |
| 89 | Q2 | 🇺🇸 | Journal of Perinatal and Neonatal Nursing | 16 | 21 |
| 91 | Q2 | 🇧🇷 | Revista Eletrônica de Enfermagem | 16 | 20 |
| 92 | Q2 | 🇨🇳 | 全科护理 (General Nursing) | 16 | 18 |
| 92 | Q2 | 🇺🇸 | Geriatric Nursing | 16 | 18 |
| 94 | Q2 | 🇺🇸 | Pain Management Nursing | 15 | 29 |
| 95 | Q2 | 🇬🇧 | Nurse Researcher | 15 | 28 |
| 96 | Q2 | 🇺🇸 | Advances in Neonatal Care | 15 | 25 |
| 97 | Q2 | 🇺🇸 | Journal for Specialists in Pediatric Nursing | 15 | 24 |
| 97 | Q2 | 🇺🇸 | Nurse Educator | 15 | 24 |
| 99 | Q2 | 🇺🇸 | Pediatric Nursing | 15 | 23 |
| 99 | Q2 | 🇨🇳 | 上海护理 (Shanghai Care) | 15 | 23 |
| 101 | Q2 | 🇺🇸 | AACN Advanced Critical Care | 15 | 22 |
| 101 | Q2 | 🇺🇸 | Applied Nursing Research | 15 | 22 |
| 101 | Q2 | 🇺🇸 | Computers, Informatics, Nursing | 15 | 22 |
| 101 | Q2 | 🇺🇸 | Journal of Child and Adolescent Psychiatric Nursing | 15 | 22 |
| 101 | Q2 | 🇺🇸 | Journal of Family Nursing | 15 | 22 |
| 101 | Q2 | 🇺🇸 | Journal of Neuroscience Nursing | 15 | 22 |
| 107 | Q2 | 🇺🇸 | Archives of Psychiatric Nursing | 15 | 21 |
| 107 | Q2 | 🇨🇳 | 国际护理学杂志 (International Journal of Nursing) | 15 | 21 |
| 107 | Q2 | 🇬🇧 | Nursing Philosophy | 15 | 21 |
| 110 | Q2 | 🇺🇸 | Journal of Psychosocial Nursing and Mental Health Services | 15 | 20 |
| 111 | Q2 | 🇬🇧 | International Journal of Palliative Nursing | 15 | 19 |
| 112 | Q2 | 🇨🇳 | 中华护理教育 (Journal of Nursing Education) | 15 | 18 |
| 113 | Q2 | 🇺🇸 | Clinical Nurse Specialist | 15 | 17 |
| 114 | Q2 | 🇺🇸 | Nursing Science Quarterly | 14 | 20 |
| 115 | Q2 | 🇺🇸 | Holistic Nursing Practice | 14 | 19 |
| 115 | Q2 | 🇬🇧 | Journal of Research in Nursing | 14 | 19 |
| 115 | Q2 | 🇺🇸 | Perspectives in Psychiatric Care | 14 | 19 |
| 118 | Q2 | 🇺🇸 | Critical Care Nursing Quarterly | 14 | 18 |
| 118 | Q2 | 🇺🇸 | Journal for nurses in staff development : JNSD : official journal of the National Nursing Staff Development Organization | 14 | 18 |
| 118 | Q2 | 🇺🇸 | Journal of School Nursing | 14 | 18 |
| 118 | Q2 | 🇨🇳 | 现代临床护理 (Modern clinical care) | 14 | 18 |
| 122 | Q2 | 🇧🇷 | Ciencia, Cuidado e Saude | 14 | 17 |
| 122 | Q2 | 🇬🇧 | Journal of Child Health Care | 14 | 17 |
| 122 | Q2 | 🇺🇸 | MedSurg Nursing | 14 | 17 |
| 122 | Q2 | 🇺🇸 | Neonatal Network | 14 | 17 |
| 122 | Q2 | 🇧🇷 | Revista Gaucha de Enfermagem | 14 | 17 |
| 127 | Q2 | 🇺🇸 | Journal of Nursing Research | 14 | 16 |
| 127 | Q2 | 🇧🇷 | Revista Enfermagem UERJ | 14 | 16 |
| 129 | Q2 | 🇩🇪 | Das Gesundheitswesen | 13 | 25 |
| 130 | Q2 | 🇺🇸 | Journal of PeriAnesthesia Nursing | 13 | 24 |
| 131 | Q2 | 🇺🇸 | Urologic Nursing | 13 | 22 |
| 131 | Q2 | 🇳🇱 | Women & Birth | 13 | 22 |
| 133 | Q2 | 🇰🇷 | Korean Academy of Nursing. Journal | 13 | 21 |
| 134 | Q2 | 🇺🇸 | Nephrology Nursing Journal | 13 | 20 |
| 135 | Q2 | 🇺🇸 | Critical Care Nursing Clinics of North America | 13 | 19 |
| 136 | Q2 | 🇺🇸 | Clinical Nursing Research | 12 | 21 |
| 137 | Q2 | 🇺🇸 | Research and Theory for Nursing Practice | 12 | 19 |
| 138 | Q2 | 🇺🇸 | Journal of Holistic Nursing | 12 | 18 |
| 138 | Q2 | 🇺🇸 | Journal of Infusion Nursing | 12 | 18 |
| 140 | Q2 | 🇳🇱 | Collegian | 12 | 17 |
| 140 | Q2 | 🇺🇸 | Nursing Clinics of North America | 12 | 17 |
| 140 | Q2 | 🇨🇦 | Nursing Leadership | 12 | 17 |
| 143 | Q2 | 🇺🇸 | AANA Journal | 12 | 16 |
| 143 | Q2 | 🇺🇸 | AAOHN Journal | 12 | 16 |
| 143 | Q2 | 🇬🇧 | International Emergency Nursing | 12 | 16 |
| 146 | Q2 | 🇺🇸 | American Psychiatric Nurses Association. Journal | 12 | 15 |
| 146 | Q2 | 🇺🇸 | Australian Critical Care | 12 | 15 |
| 146 | Q2 | 🇧🇷 | Cogitare Enfermagem | 12 | 15 |
| 146 | Q2 | 🇨🇳 | 临床护理杂志 (Journal of Clinical Nursing) | 12 | 15 |
| 146 | Q2 | 🇺🇸 | Nursing Management | 12 | 15 |
| 151 | Q2 | 🇺🇸 | Rehabilitation Nursing | 12 | 14 |
| 152 | Q2 | 🇨🇦 | Canadian Journal of Nursing Research | 12 | 13 |
| 153 | Q2 | 🇦🇺 | International Journal of Evidence-Based Healthcare | 11 | 21 |
| 154 | Q2 | 🇺🇸 | Dimensions of Critical Care Nursing | 11 | 18 |
| 155 | Q2 | 🇺🇸 | Journal of Hospice and Palliative Nursing | 11 | 15 |
| 156 | Q2 | 🇫🇷 | Journal of Renal Care | 11 | 14 |
| 157 | Q2 | 🇨🇦 | Canadian Nurse | 11 | 12 |
| 158 | Q2 | 🇺🇸 | Journal of Cultural Diversity | 10 | 17 |
| 159 | Q2 | 🇺🇸 | Clinical Simulation in Nursing | 10 | 16 |
| 159 | Q2 | 🇺🇸 | Professional Case Management | 10 | 16 |
| 161 | Q2 | 🇺🇸 | Teaching and Learning in Nursing | 10 | 15 |
| 161 | Q2 | 🇨🇳 | 天津护理 (Tianjin care) | 10 | 15 |
| 163 | Q2 | 🇺🇸 | Gastroenterology nursing : the official journal of the Society of Gastroenterology Nurses and Associates | 10 | 14 |
| 163 | Q2 | 🇬🇧 | International Journal of Older People Nursing (Online) | 10 | 14 |
| 163 | Q2 | 🇺🇸 | Nurse Practitioner | 10 | 14 |
| 163 | Q2 | 🇺🇸 | Plastic Surgical Nursing | 10 | 14 |
| 167 | Q2 | 🇬🇧 | British Journal of Community Nursing | 10 | 13 |
| 167 | Q2 | 🇬🇧 | British Journal of Midwifery | 10 | 13 |
| 167 | Q2 | 🇬🇧 | Nursing Times | 10 | 13 |
| 167 | Q2 | 🇺🇸 | Research in gerontological nursing | 10 | 13 |
| 171 | Q3 | 🇺🇸 | Home Healthcare Nurse | 10 | 11 |
| 172 | Q3 | 🇨🇱 | Ciencia y Enfermería | 9 | 15 |
| 172 | Q3 | 🇺🇸 | Journal of Trauma Nursing | 9 | 15 |
| 172 | Q3 | 🇺🇸 | Newborn and Infant Nursing Reviews | 9 | 15 |
| 172 | Q3 | 🇬🇧 | Nursing Older People | 9 | 15 |
| 176 | Q3 | 🇬🇧 | Journal of Neonatal Nursing | 9 | 14 |
| 177 | Q3 | 🇧🇷 | Revista da Rede de Enfermagem do Nordeste | 9 | 12 |
| 178 | Q3 | 🇪🇸 | Index de Enfermería | 9 | 11 |
| 179 | Q3 | 🇺🇸 | Journal of Forensic Nursing | 8 | 14 |
| 180 | Q3 | 🇺🇸 | Nursing for Women's Health | 8 | 13 |
| 181 | Q3 | 🇨🇴 | Aquichán | 8 | 12 |
| 181 | Q3 | 🇰🇷 | Korean Journal of Women Health Nursing | 8 | 12 |
| 183 | Q3 | 🇺🇸 | Alzheimer's Care Today | 8 | 11 |
| 183 | Q3 | 🇺🇸 | Journal for Nurse Practitioners | 8 | 11 |
| 183 | Q3 | 🇰🇷 | Korean Academy of Child Health Nursing. Journal | 8 | 11 |
| 186 | Q3 | 🇬🇧 | Australasian Emergency Nursing Journal | 8 | 10 |
| 186 | Q3 | 🇺🇸 | Bariatric Nursing & Surgical Patient Care | 8 | 10 |
| 186 | Q3 | 🇨🇳 | 中华现代护理学杂志 (Journal of Modern Nursing) | 8 | 10 |
| 186 | Q3 | 🇬🇧 | Journal of Perioperative Practice | 8 | 10 |
| 190 | Q3 | 🇺🇸 | ABNF Journal | 8 | 9 |
| 190 | Q3 | 🇬🇧 | Community Practitioner | 8 | 9 |
| 190 | Q3 | 🇬🇧 | Nursing Management – UK | 8 | 9 |
| 193 | Q3 | 🇺🇸 | Journal of Vascular Nursing | 7 | 15 |
| 194 | Q3 | 🇪🇸 | Gerokomos | 7 | 12 |
| 195 | Q3 | 🇺🇸 | Advanced Emergency Nursing Journal | 7 | 10 |
| 195 | Q3 | 🇺🇸 | Hispanic Health Care International | 7 | 10 |
| 195 | Q3 | 🇧🇷 | Online Brazilian Journal of Nursing | 7 | 10 |
| 195 | Q3 | 🇫🇷 | Pratiques et organisation des soins | 7 | 10 |
| 199 | Q3 | 🇨🇦 | Canadian Oncology Nursing Journal | 7 | 9 |
| 199 | Q3 | 🇪🇸 | Enfermería Clínica | 7 | 9 |
| 199 | Q3 | 🇰🇷 | Korean Journal of Adult Nursing | 7 | 9 |
| 202 | Q3 | 🇨🇦 | CANNT Journal | 7 | 8 |
| 202 | Q3 | 🇬🇧 | Journal of Family Health Care | 7 | 8 |
| 202 | Q3 | 🇨🇺 | Revista Cubana de Enfermería | 7 | 8 |
| 205 | Q3 | 🇬🇧 | Asian Nursing Research | 7 | 7 |
| 206 | Q3 | 🇬🇧 | Evidence - Based Nursing | 6 | 20 |
| 207 | Q3 | 🇺🇸 | Perioperative Nursing Clinics | 6 | 15 |
| 208 | Q3 | 🇺🇸 | Creative Nursing | 6 | 11 |
| 209 | Q3 | 🇨🇴 | Avances en Enfermería | 6 | 10 |
| 210 | Q3 | 🇪🇸 | Enfermería Global | 6 | 9 |
| 210 | Q3 | 🇬🇧 | Nurse Prescribing | 6 | 9 |
| 210 | Q3 | 🇧🇷 | Revista Mineira de Enfermagem | 6 | 9 |
| 213 | Q3 | 🇺🇸 | International Journal for Human Caring | 6 | 8 |
| 213 | Q3 | 🇬🇧 | Journal of Addictions Nursing | 6 | 8 |
| 213 | Q3 | 🇺🇸 | Journal of Radiology Nursing | 6 | 8 |
| 216 | Q3 | 🇬🇧 | Emergency Nurse | 6 | 7 |
| 216 | Q3 | 🇪🇸 | Enfermería Intensiva | 6 | 7 |
| 216 | Q3 | 🇺🇸 | Nurse Leader | 6 | 7 |
| 216 | Q3 | 🇪🇸 | Revista de la Sociedad Española de Enfermería Nefrológica | 6 | 7 |
| 220 | Q3 | 🇵🇱 | Problemy Pielęgniarstwa | 6 | 6 |
| 220 | Q3 | 🇧🇷 | UFPE Revista de Enfermagem on Line/ Journal of Nursing UFPE | 6 | 6 |
| 222 | Q3 | 🇺🇸 | Nursing History Review | 5 | 11 |
| 223 | Q3 | 🇬🇧 | European Diabetes Nursing | 5 | 9 |
| 224 | Q3 | 🇪🇸 | Metas de Enfermería | 5 | 8 |
| 225 | Q3 | 🇪🇸 | Revista Rol de Enfermería | 5 | 7 |
| 226 | Q3 | 🇦🇺 | Australian Nursing Journal | 5 | 6 |
| 226 | Q3 | 🇮🇷 | Iranian Journal of Nursing and Midwifery Research | 5 | 6 |
| 226 | Q3 | 🇨🇭 | Pflege | 5 | 6 |
| 229 | Q3 | 🇰🇷 | Journal of Korean Academy of Fundamentals of Nursing | 5 | 5 |
| 230 | Q3 | 🇺🇸 | Insight: The Journal of the American Society of Ophthalmic Registered Nurses | 4 | 12 |
| 231 | Q3 | 🇨🇴 | Investigación y Educación en Enfermería | 4 | 9 |
| 232 | Q3 | 🇬🇧 | British Journal of Anaesthetic & Recovery Nursing | 4 | 7 |
| 232 | Q3 | 🇬🇧 | Journal of Community Nursing | 4 | 7 |
| 232 | Q3 | 🇺🇸 | ONS Connect | 4 | 7 |
| 235 | Q3 | 🇬🇧 | British Journal of Neuroscience Nursing | 4 | 6 |
| 235 | Q3 | 🇪🇸 | Cultura de los Cuidados | 4 | 6 |
| 235 | Q3 | 🇬🇧 | International Journal of Urological Nursing | 4 | 6 |
| 238 | Q3 | 🇬🇧 | Cancer Nursing Practice | 4 | 5 |
| 238 | Q3 | 🇮🇷 | Hayat | 4 | 5 |

| Rank | Quartile | | Journal | H | H5 |
|---|---|---|---|---|---|
| 238 | Q3 | 🇬🇧 | International Journal of Orthopaedic and Trauma Nursing | 4 | 5 |
| 238 | Q3 | 🇰🇷 | 아동간호학회지 (Journal of Child Health Nursing) | 4 | 5 |
| 238 | Q3 | 🇬🇧 | Mental Health Today | 4 | 5 |
| 238 | Q3 | 🇬🇧 | Practice Nursing | 4 | 5 |
| 238 | Q3 | 🇨🇳 | 護理雜誌 (The Journal of Nursing) | 4 | 5 |
| 245 | Q3 | 🇺🇸 | Dermatology Nurses' Association. Journal | 4 | 4 |
| 245 | Q3 | 🇬🇧 | Gastrointestinal Nursing | 4 | 4 |
| 245 | Q3 | 🇺🇸 | Journal of Christian Nursing | 4 | 4 |
| 245 | Q3 | 🇺🇸 | Journal of the Dermatology Nurses' Association | 4 | 4 |
| 245 | Q3 | 🇬🇧 | Nursing & Residential Care | 4 | 4 |
| 245 | Q3 | 🇮🇳 | Nursing Journal of India | 4 | 4 |
| 245 | Q3 | 🇬🇧 | The Practising Midwife | 4 | 4 |
| 252 | Q3 | 🇫🇷 | Recherche en Soins Infirmiers | 3 | 22 |
| 253 | Q4 | 🇨🇴 | Investigación en Enfermería: Imagen y Desarrollo | 3 | 19 |
| 253 | Q4 | 🇮🇹 | Professioni Infermieristiche | 3 | 19 |
| 255 | Q4 | 🇫🇷 | Soins: La Revue de Reference Infirmiere | 3 | 10 |
| 256 | Q4 | 🇺🇸 | Beginnings | 3 | 9 |
| 257 | Q4 | 🇬🇧 | African Journal of Midwifery & Women's Health | 3 | 8 |
| 258 | Q4 | 🇺🇸 | NASN School Nurse | 3 | 8 |
| 259 | Q4 | 🇰🇷 | 여성건강간호학회지 (Journal of Women's Health Nursing) | 3 | 7 |
| 259 | Q4 | 🇨🇦 | Perspective Infirmiere | 3 | 7 |
| 259 | Q4 | 🇫🇷 | Revue Sage - Femme | 3 | 7 |
| 262 | Q4 | 🇬🇧 | British Journal of Cardiac Nursing | 3 | 6 |
| 262 | Q4 | 🇬🇧 | British Journal of School Nursing | 3 | 6 |
| 262 | Q4 | 🇪🇸 | Matronas Profesión | 3 | 6 |
| 262 | Q4 | 🇩🇪 | Pflegezeitschrift | 3 | 6 |
| 262 | Q4 | 🇧🇷 | Revista de Pesquisa: Cuidado e Fundamental (Online) | 3 | 6 |
| 262 | Q4 | 🇫🇷 | Soins Gerontologie | 3 | 6 |
| 268 | Q4 | 🇺🇸 | American Nurse | 3 | 5 |
| 268 | Q4 | 🇮🇹 | Assistenza Infermieristica e Ricerca | 3 | 5 |
| 268 | Q4 | 🇬🇧 | British Journal of Healthcare Assistants | 3 | 5 |
| 268 | Q4 | 🇰🇷 | 한국보건간호학회지 (Korea Health Nursing) | 3 | 5 |
| 268 | Q4 | 🇺🇸 | School Nurse News | 3 | 5 |
| 273 | Q4 | 🇺🇸 | Imprint | 3 | 4 |
| 273 | Q4 | 🇬🇧 | Journal of Diabetes Nursing | 3 | 4 |
| 273 | Q4 | 🇰🇷 | 간호행정학회지 (Journal of Nursing Administration) | 3 | 4 |
| 273 | Q4 | 🇰🇷 | 정신간호학회지 (Journal of Psychiatric and Mental Health Nursing) | 3 | 4 |
| 273 | Q4 | 🇺🇸 | Midwifery Today | 3 | 4 |
| 273 | Q4 | 🇺🇸 | ORL-Head & Neck Nursing | 3 | 4 |
| 273 | Q4 | 🇿🇦 | Professional Nursing Today | 3 | 4 |
| 273 | Q4 | 🇵🇹 | Referência | 3 | 4 |
| 281 | Q4 | 🇹🇷 | Anadolou Hemsirelik ve Saglik Bilimleri Dergisi | 3 | 3 |
| 281 | Q4 | 🇬🇧 | Journal of Renal Nursing | 3 | 3 |
| 283 | Q4 | 🇪🇸 | Enfermería Integral | 2 | 6 |
| 284 | Q4 | 🇧🇷 | Revista Rede de Cuidados em Saúde | 2 | 4 |
| 285 | Q4 | 🇨🇦 | Alberta RN | 2 | 3 |
| 285 | Q4 | 🇨🇳 | 志為護理-慈濟護理雜誌 (Chi Care - Nursing magazine) | 2 | 3 |
| 285 | Q4 | 🇰🇷 | 임상간호연구 (Clinical Nursing Research) | 2 | 3 |
| 285 | Q4 | 🇪🇸 | Hygia de enfermería: revista científica del colegio | 2 | 3 |
| 285 | Q4 | 🇮🇷 | Journal of Nursing & Midwifery | 2 | 3 |
| 285 | Q4 | 🇰🇷 | 산업간호학회지 (Industrial Nursing) | 2 | 3 |
| 285 | Q4 | 🇯🇵 | 医療看護研究 (Medical nursing research) | 2 | 3 |
| 285 | Q4 | 🇯🇵 | 看護教育 (Nursing education) | 2 | 3 |
| 285 | Q4 | 🇯🇵 | 看護研究 (Nursing research) | 2 | 3 |
| 285 | Q4 | 🇺🇸 | Pennsylvania Nurse | 2 | 3 |
| 285 | Q4 | 🇩🇪 | Psychiatrische Pflege Heute | 2 | 3 |
| 296 | Q4 | 🇪🇸 | Ene | 2 | 2 |
| 296 | Q4 | 🇯🇵 | 老年看護学: 日本老年看護学会誌: Journal of Japan Academy of Gerontological Nursing | 2 | 2 |
| 296 | Q4 | 🇯🇵 | 日本がん看護学会誌 (Japan Cancer Nursing Journal) | 2 | 2 |
| 296 | Q4 | 🇯🇵 | 日本看護科学会誌 (Japan Journal of Nursing Science) | 2 | 2 |
| 296 | Q4 | 🇯🇵 | 日本看護学会論文集, 看護教育 (Japan Journal of Nursing, nursing education) | 2 | 2 |
| 296 | Q4 | 🇯🇵 | 日本看護研究学会雑誌 (Japan Society of Nursing Research Journal) | 2 | 2 |
| 296 | Q4 | 🇯🇵 | 日本小児看護学会誌 (Japanese Society of Child Health Nursing Journal) | 2 | 2 |
| 296 | Q4 | 🇨🇳 | 榮總護理 (Journal of Advanced Nursing) | 2 | 2 |
| 296 | Q4 | 🇰🇷 | 종양간호학회지 (Journal of Oncology Nursing) | 2 | 2 |
| 296 | Q4 | 🇬🇧 | Nursing in Practice | 2 | 2 |
| 296 | Q4 | 🇯🇵 | 看護学雑誌 (Nursing Journal) | 2 | 2 |
| 296 | Q4 | 🇰🇷 | 기본간호학회지 (Nursing Journal) | 2 | 2 |
| 296 | Q4 | 🇯🇵 | 山梨大学看護学会誌 (Nursing Journal of University of Yamanashi) | 2 | 2 |
| 296 | Q4 | 🇯🇵 | 看護管理 (Nursing Management) | 2 | 2 |
| 296 | Q4 | 🇨🇳 | 長庚護理 (Nursing Research) | 2 | 2 |
| 296 | Q4 | 🇪🇸 | REDUCA (Enfermería, Fisioterapia y Podología) | 2 | 2 |
| 296 | Q4 | 🇫🇷 | Revue de l'Infirmiere | 2 | 2 |
| 296 | Q4 | 🇫🇷 | Soins Cadres | 2 | 2 |
| 296 | Q4 | 🇫🇷 | Soins Pediatrie - Puericulture | 2 | 2 |
| 296 | Q4 | 🇫🇷 | Soins Psychiatrie | 2 | 2 |
| 296 | Q4 | 🇳🇱 | Tijdschrift voor verpleeghuisgeneeskunde | 2 | 2 |
| 317 | Q4 | 🇫🇷 | Droit, Déontologie & Soin | 1 | 7 |
| 318 | Q4 | 🇯🇵 | 日本看護学会論文集, 看護総合 (Japan Journal of Nursing, Nursing Research) | 1 | 6 |
| 319 | Q4 | 🇯🇵 | 日本看護学会論文集, 成人看護(Japan Journal of Nursing, Adult Nursing) | 1 | 4 |
| 319 | Q4 | 🇺🇸 | Journal of Community Health Nursing | 1 | 4 |
| 319 | Q4 | 🇰🇷 | 지역사회간호학회지 (Journal of Community Health Nursing) | 1 | 4 |
| 319 | Q4 | 🇰🇷 | 성인간호학회지 (Nursing) | 1 | 4 |
| 323 | Q4 | 🇪🇸 | Ágora de Enfermería | 1 | 3 |
| 323 | Q4 | 🇨🇭 | Krankenpflege. Soins infirmiers | 1 | 3 |
| 323 | Q4 | 🇳🇱 | TvZ: Tijdschrift voor Verpleegkundigen | 1 | 3 |
| 326 | Q4 | 🇯🇵 | 千葉看護学会会誌 (Chiba Society of Nursing) | 1 | 2 |
| 326 | Q4 | 🇯🇵 | 小児看護 (Child Health Nursing) | 1 | 2 |
| 326 | Q4 | 🇯🇵 | 日本看護技術学会誌 (Japan Journal of Nursing Technology) | 1 | 2 |
| 326 | Q4 | 🇯🇵 | 日本看護学教育学会誌 (Japan Nursing Education Journal) | 1 | 2 |
| 326 | Q4 | 🇯🇵 | 看護 (Nursing) | 1 | 2 |
| 326 | Q4 | 🇰🇷 | 대한기초간호자연과학회지 (Nursing Journal of nature and the basis for) | 1 | 2 |
| 326 | Q4 | 🇯🇵 | 看護と情報: 看護図書館協議会会誌 (Nursing Library Association Journal: Information and nursing) | 1 | 2 |

| 326 | Q4 | | Queensland Nurse | 1 | 2 |
| 334 | Q4 | | International Nursing Review en Español | 1 | 1 |
| 334 | Q4 | | 領導護理 (Nursing Leadership) | 1 | 1 |
| 334 | Q4 | | Meditsinski Pregled. Sestrinsko Delo | 1 | 1 |
| 334 | Q4 | | Verpleegkunde | 1 | 1 |

## REFERENCES

For more information about the use of Google Scholar as a source for bibliometric evaluation of journals, check the following studies performed by EC3 research group: